\documentclass[12pt,preprint]{aastex}

\slugcomment{Accepted by AJ on 2016 April 12}

\shorttitle{Intermediate-Mass Black Holes in Globular Clusters}
\shortauthors{Wrobel, Miller-Jones, \& Middleton}

\begin{document}

\title{A Very Large Array Search for Intermediate-Mass Black Holes in
  Globular Clusters in M81}

\author{
  J. M. Wrobel\altaffilmark{1,}\altaffilmark{2},
  J. C. A. Miller-Jones\altaffilmark{3}, and
  M. J. Middleton\altaffilmark{4}}

\altaffiltext{1}{National Radio Astronomy Observatory, P.O. Box O,
  Socorro, NM 87801, USA; jwrobel@nrao.edu}

\altaffiltext{2}{National Science Foundation, 4201 Wilson Boulevard,
  Arlington, VA 22230, USA; jwrobel@nsf.gov}

\altaffiltext{3}{International Centre for Radio Astronomy Research,
  Curtin University, GPO Box U1987, Perth, WA 6845, Australia;
  james.miller-jones@curtin.edu.au}

\altaffiltext{4}{Institute of Astronomy, University of Cambridge, 
Madingley Road, Cambridge CB3 0HA, UK; mjm@ast.cam.ac.uk}

\begin{abstract}
  Nantais et al.\ used the {\em Hubble Space Telescope\/} to localize
  probable globular clusters (GCs) in M81, a spiral galaxy at a
  distance of 3.63 Mpc.  Theory predicts that GCs can host
  intermediate-mass black holes (IMBHs) with masses $M_{\rm BH} \sim
  100 - 100,000~M_\odot$.  Finding IMBHs in GCs could validate a
  formation channel for seed BHs in the early universe, bolster
  gravitational-wave predictions for space missions, and test scaling
  relations between stellar systems and the central BHs they host.  We
  used the NRAO Karl G.\ Jansky Very Large Array (VLA) to search for
  the radiative signatures of IMBH accretion from 206 probable GCs in
  a mosaic of M81.  The observing wavelength was 5.5 cm and the
  spatial resolution was 1.5\arcsec\ (26.4 pc).  None of the
  individual GCs are detected, nor are weighted-mean image stacks of
  the 206 GCs and the 49 massive GCs with stellar masses $M_\star
  \gtrsim 200,000~M_\odot$.  We apply a semi-empirical model to
  predict the mass of an IMBH that, if undergoing accretion in the
  long-lived hard X-ray state, is consistent with a given radio
  luminosity.  The 3$\sigma$ radio-luminosity upper limits correspond
  to IMBH masses of $\overline{M_{\rm BH}({\rm all})} <
  42,000~M_\odot$ for the all-cluster stack and $\overline{M_{\rm
      BH}({\rm massive})} < 51,000~M_\odot$ for the massive-cluster
  stack.  We also apply the empirical fundamental-plane relation to
  two X-ray-detected clusters, finding that their individual IMBH
  masses at 95\% confidence are $M_{\rm BH} < 99,000~M_\odot$ and
  $M_{\rm BH} < 15,000~M_\odot$.  Finally, no analog of HLX-1, a
  strong IMBH candidate in an extragalactic star cluster, occurs in
  any individual GC in M81.  This underscores the uniqueness or rarity
  of the HLX-1 phenomenon.
\end{abstract}


\section{Motivation}\label{motivation}

Intermediate-mass black holes (IMBHs) are thought to occupy the mass
gap between the well-established stellar-mass BHs with $M_{\rm BH} <
100~M_\odot$ \citep[e.g.,][]{abb16a,abb16b,cor16,tet16} and the
well-studied supermassive BHs with $M_{\rm BH} \gtrsim 10^6~M_\odot$
\citep[e.g.,][]{kor13}.  Finding IMBHs in the local universe can
provide important insight into formation channels for seed BHs in the
early universe \citep[for reviews, see][]{gre12,vol12,nat14}.  Here,
we focus on theoretical predictions that globular clusters (GCs) could
host IMBHs formed via dynamical processes stemming from the clusters'
closely packed stars \citep[e.g.,][]{mil02,gur04,por04,gie15}.  In
addition, if IMBHs exist in GCs, inspirals involving them and
stellar-mass BHs could be years-long sources of gravitational waves
for a LISA-like mission \citep{kon13}.  Such IMBHs could also be used
to test scaling relations between central BHs and their stellar system
hosts, thus informing the debate about whether or not these entities
co-evolve \citep[e.g.,][]{xia11,jia11,kor13,gra16}.

A key science driver for future near-infrared, ground-based telescopes
is to measure, at a distance of 10 Mpc, a BH mass as low as $M_{\rm
  BH} \sim 10^5~M_\odot$ by spatially resolving its sphere of
influence in its host stellar system \citep{do14}.  It is expected
that this will yield a robust inventory of IMBHs in GCs in the local
universe, but these next-generation facilities are many years off.
Moreover, although GCs in the Local Group have been targeted in
sphere-of-influence studies, all these studies are contentious
\citep[][and references therein]{str12a}.

Given these controversies and long waits for future facilities, we
have begun an independent investigation, namely searching for radio
signatures of accretion onto putative IMBHs in extragalactic GCs
\citep{wro15}.  By analogy with stellar-mass BHs \citep[reviewed
  by][]{fen12}, one expects that an IMBH will spend more time in the
hard X-ray state - including quiesence - associated with a low
accretion rate onto the BH, than in the soft X-ray state associated
with a high accretion rate.  In the typical case of only a few radio
observations, it is likely that they will sample the steady radio
emission characteristic of the hard X-ray state, as opposed to the
flaring radio emssion associated with a transition from the hard X-ray
state to the soft X-ray state.  These concepts, first laid out by
\citet{mac04}, lead to the following three approaches:

\begin{description}
\item [(1)] Detect radio emission like that from HLX-1 in its hard
  X-ray state, where HLX-1 is a strong IMBH candidate in an
  extragalactic star cluster with a stellar mass of $M_\star \sim
  10^{5-6}~M_\odot$ \citep{cse15}.
\item [(2)] Use the empirical fundamental-plane regression for the
  hard X-ray state, plus observations of X-ray and radio luminosities,
  to estimate an IMBH mass \citep{mer03,fal04,plo12}.
\item [(3)] Use a conservative, semi-empirical model to predict the
  mass of an IMBH that, if experiencing Bondi accretion in the hard
  X-ray state, would be consistent with the observed radio luminosity
  \citep{mac08,mac10,str12a}.
\end{description}

Our radio imaging of NGC\,1023, an early-type galaxy about 11 Mpc
away, detected none of its 337 candidate star clusters \citep{wro15}.
The approaches above then led to the following inferences: (1) No
HLX-1 analogs suggests that phenomenon is very rare or is fuelled from
gas related to its cluster's relatively young stars. (2) To be able to
reach the regime of IMBH masses, deeper X-ray and radio surveys that
detect candidate clusters at lower luminosities are needed.
Importantly, the mass term in the fundamental plane implies that, in
the hard-X-ray state, the radio emission from IMBHs should be much
brighter than that from stellar-mass BHs.  This makes radio detections
effective at filtering out contamination from the X-ray-emitting
stellar-mass BHs often found in star clusters. (3) The
radio-luminosity upper limit for a stack of the 20 most massive
clusters corresponds to a mean 3$\sigma$ IMBH mass of
$\overline{M_{\rm BH}({\rm massive})} < 2.3\times 10^5~M_\odot$ and a
BH mass fraction $\overline{M_{\rm BH}({\rm
    massive})}/\overline{M_\star({\rm massive})} < 0.16$.

The inferences from our NGC\,1023 study can be improved upon by
obtaining longer exposures on star clusters in closer galaxies.  We
thus turn to \object[M81]{M81}, a spiral galaxy at a distance of
3.63$\pm$0.34 Mpc \citep[1\arcsec\, = 17.6 pc;][]{fre94} that is
estimated to have 210$\pm$30 GCs in total \citep{per95}.  Several
studies have used the {\em Hubble Space Telescope (HST)\/} to localize
candidate GCs in M81 \citep{cha01,cha04,nan10,san10,nan11}.  Here, we
focus on the \citet{nan11} study because it leveraged Advanced Camera
for Surveys (ACS) mosaics with spectroscopy to identify probable GCs
in M81.  \citet{nan11} performed aperture photometry and profile
fitting on 419 candidate GCs, and highlight 214\footnote{Nantais et
  al. (2011) tabulate this number but mistakenly quote 221 in their
  text.} as being probable GCs because they were spectroscopically
confirmed (85) or were good candidates (129) that shared the color and
size ranges of the spectroscopically confirmed GCs.  The V-band
(F606W) magnitudes from \citet{nan11} and that band's mass-to-light
ratio for GCs from \citet{har10} imply that the stellar masses of the
214 probable GCs range from $M_\star \sim 1 \times 10^4~M_\odot$ to
$M_\star \sim 8 \times 10^6~M_\odot$.

In this paper, we use the NRAO\footnote{The National Radio Astronomy
  Observatory is a facility of the National Science Foundation,
  operated under cooperative agreement by Associated Universities,
  Inc.} Karl G.\ Jansky Very Large Array \citep[VLA;][]{per11} to
search for radio emission from these probable GCs in M81.  We describe
our new VLA results in \S~\ref{imaging}.  No individual probable GC is
detected.  The implications of these nondetections are then explored
regarding HLX-1 analogs (\S~\ref{hlx}), X-ray detected clusters
(\S~\ref{x-ray}), and semi-empirical model predictions
(\S~\ref{model}).  We close in \S~\ref{sumcon} with a summary and
conclusions.

\section{Imaging}\label{imaging}

We observed M81 under proposal code 13B-138 (PI M.\ Middleton) using
the VLA in its B configuration at a central frequency of 5.5 GHz,
corresponding to a wavelength of 5.5 cm.  Owing to the large angular
size of the galaxy, we used four pointing centers separated by
4\farcm5 to provide relatively uniform sensitivity to the central
regions of the galaxy.  The pointing centers' locations and
observation dates appear in Table 1.

Each VLA pointing was observed for one hour, giving 41 m of time on
target, using a correlator integration time of 3 s.  The target
elevation lay between 23$\arcdeg$ and 32$\arcdeg$ in all cases.  The
observing bandwidth was split into two 1024-MHz basebands centred at
5.0 and 6.0 GHz, each comprised of eight spectral windows of width
128 MHz, each of which was split into sixty-four 2-MHz channels.  We
used 3C\,147 to set the amplitude scale to an estimated accuracy of
about 3\%, and J1048+7143 as the secondary calibrator to derive the
atmospheric and instrumental complex gains on a per-antenna basis.
The position assumed for J1048+7143 was $\alpha(J2000) = 10^{h} 48^{m}
27\fs6199$ and $\delta(J2000) = 71\arcdeg 43\arcmin 35\farcs938$ with
one-dimensional errors at 1$\sigma$ of 2 mas.  Given the observing
strategies, the one-dimensional astrometric error at 1$\sigma$ is
estimated to be 0\farcs1.

We processed the data using standard procedures within the Common
Astronomy Software Application \citep[CASA;][]{mcm07} release 4.1.0.
Following external gain calibration, the Stokes $I\/$ data for each
pointing were imaged with the CASA task {\tt clean} using a robustness
parameter of 0.5 to obtain the best compromise among sensitivity,
spatial resolution and sidelobe suppression; an nterms parameter of 2
to accommodate the large fractional bandwidth; and the gridmode
parameter set to ``widefield'' and the wprojplanes parameter set to
128 to correct for the effects of non-coplanar baselines.  The images
were dominated by the emission from the low-luminosity active galactic
nucleus (LLAGN) in M81, and required self-calibration, initially in
phase, and then in amplitude and phase, down to a timescale as short
as the correlator integration time of 3 s.  The final, self-calibrated
images of all four quadrants were restored with the same circular
Gaussian restoring beam of FWHM of 1\farcs5, and then mosaicked
together, cutting off the response of the primary beam at the 10\%
level.  In the mosaic, the LLAGN is point-like and has a flux density
of 81$\pm$3 mJy.  Figure 1 shows the geometry of the VLA mosaic
overlaid on a $K_s$ image of M81 \citep{jar03} retrieved from NED.

\begin{figure}[t]
\plotone{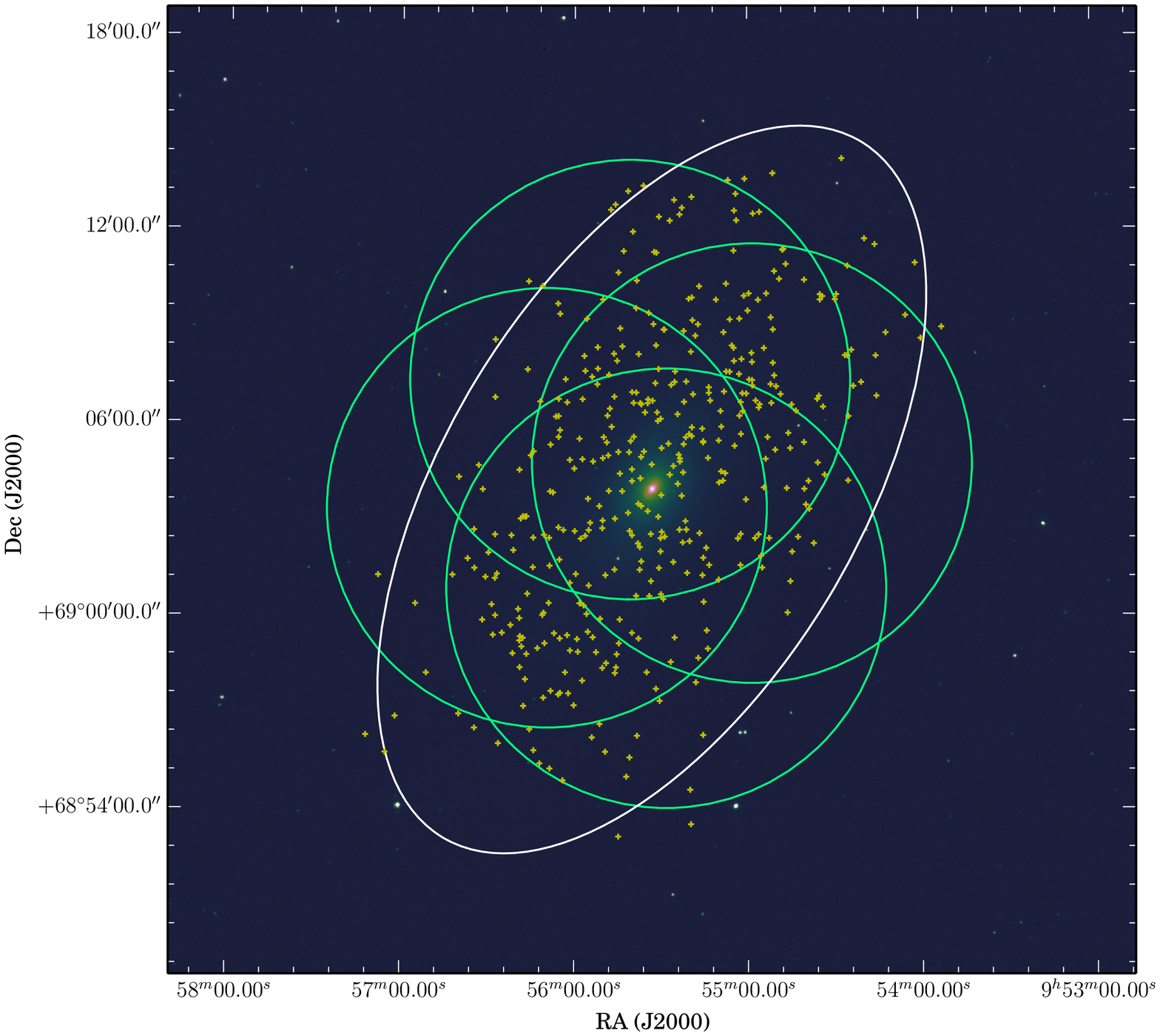}
\caption{Four overlapping green circles indicate the VLA mosaic region
  for M81 at 5.5 GHz.  Each circle is located at one of the VLA
  pointing centers (Table 1) and has a diameter of 13\farcm6 (14.4
  kpc), where the response of the VLA primary beam falls to the 10\%
  level.  The colored shading shows the surface brightness of the
  inner galaxy in the $K_s$ band, while the white ellipse conveys the
  extent of the galaxy in that band as deduced by Jarrett et
  al.\ (2003).  The yellow plus signs mark the locations of the 419
  candidate GCs from Nantais et al.\ (2011).}\label{f1}
\end{figure}

For each of the 419 candidate GCs, task {\tt subim} in the 2016
December 31 release of NRAO's Astronomical Image Processing System
\citep[AIPS;][]{gre03} was used to form a cutout spanning 30$\arcsec$
and centered on the optical position \citep{nan11}.  Astrometric
comparions between the ACS images and ground-based surveys implied
conservative one-dimensional errors at 1$\sigma$ of 0\farcs2
\citep{nan11}.  Thirteen candidate GCs lay beyond the boundary of the
VLA mosaic.  This left 406 candidates for analysis, split as 206
probable GCs and 200 improbable GCs.  The term ``improbable'' refers
to candidate GCs that \citet{nan11} judged to be background galaxies
or non-GC objects like young clusters and HII regions, or confirmed as
galaxies or non-GC objects via spectroscopy.

Figure 2 shows the VLA cutouts for the 206 probable GCs.  The
1$\sigma$ rms noise level among the cutouts varies from 4.3 to
51~$\mu$Jy~beam$^{-1}$, a range expected given the primary beam cutoff
invoked.  An individual one-tail detection threshold of 3$\sigma$ is
adopted to minimize the risk of a false-positive detection of one or
more targets when examining the ensembles of 206 probable GCs and 200
improbable GCs \citep{wal03}.

\begin{figure}[t]
\plotone{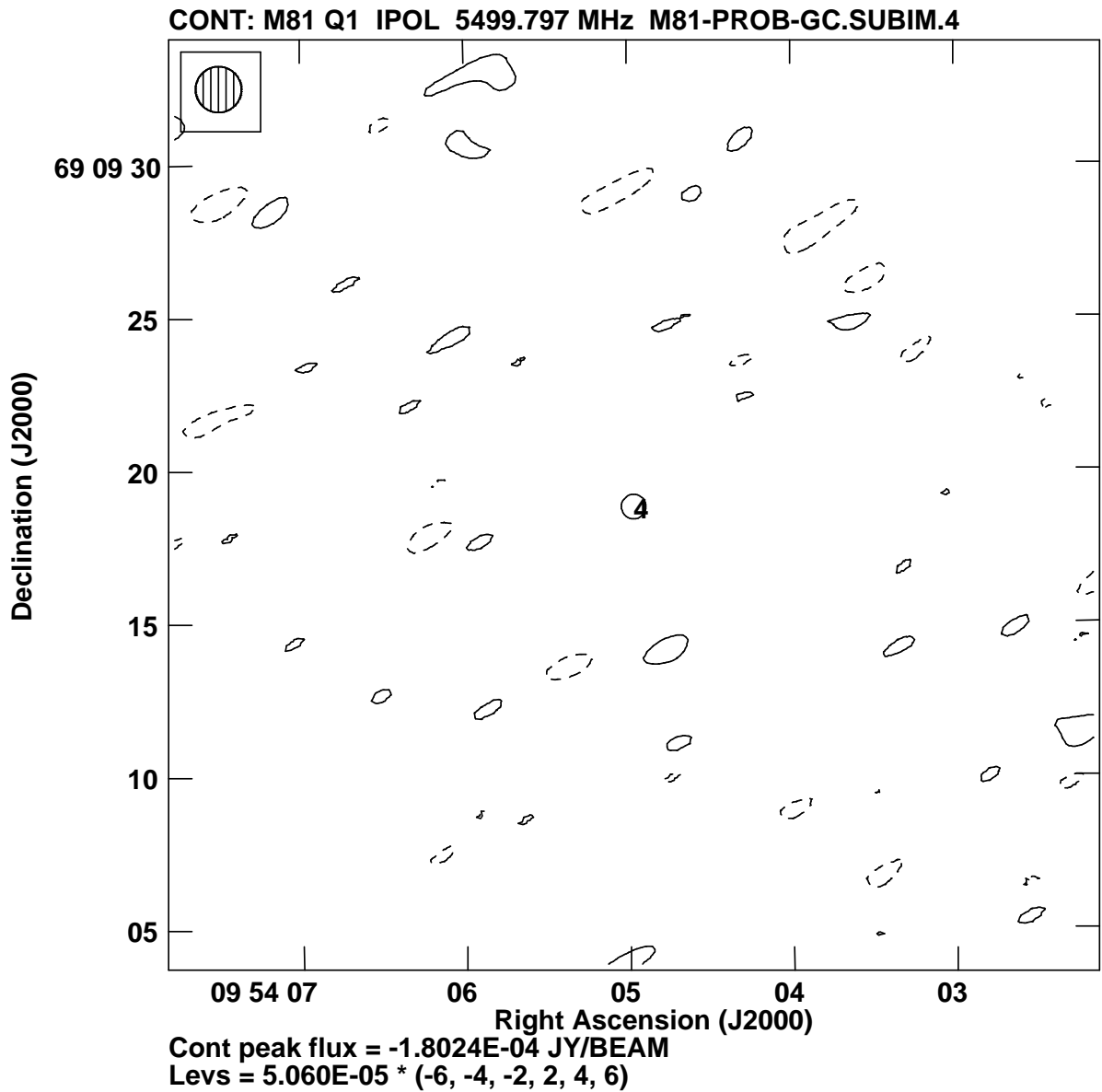}
\caption{VLA cutout of the Stokes $I\/$ emission at 5.5 GHz centered
  on the optical position of a probable GC in M81.  The cutout spans
  30$\arcsec$ (1.6 kpc) per coordinate.  The diameter of the hatched
  circle in the north-east corner, 1\farcs5 (26.4 pc), is the FWHM of
  the VLA synthesized beam.  Contours are at -6, -4, -2, 2, 4, and 6
  times the 1$\sigma$ rms noise shown in the legend in units of
  Jy~beam$^{-1}$.  Linearly-spaced contours are chosen to convey noise
  levels at a glance, with dashed lines showing negative contours and
  solid lines showing positive ones.  The central circle of diameter
  0\farcs8 (32 pc) shows the cluster's optical positional uncertainty
  at 90\% confidence and is labelled with its ID from Nantais et
  al.\ (2011).  The VLA photometry seeks evidence for the accretion
  signature of a point-like IMBH in the cluster's center.  The cluster
  is not detected above the 3$\sigma$ level.  Figures 2.1 $-$ 2.206
  are available, ordered by increasing ID, in the online version of
  the Journal.  Some cutouts show additional candidate GCs offset from
  the central probable GC.}\label{f2}
\end{figure}

Only one of the 200 improbable GCs, ID 146 from \citet{nan11}, was
detected with enough significance to serve as a crosscheck of the
radio and optical astrometry.  The VLA cutout of ID 146 appears in
Figure 3.  The radio emission is somewhat resolved but dominantly
compact.  The radio peak has a total offset of $0\farcs2$ when
compared to the optical position reported by \citet{nan11}.  Assuming
that the peaks should physcially coincide, such an offset is
consistent with the error estimates cited above for the radio and
optical data.  Moreover, finding resolved radio emission from ID 146
is consistent with its classification as an improbable GC: if it was a
probable GC hosting an IMBH in the low-hard state, any radio emission
should be spatially unresolved.

\begin{figure}[t]
\plotone{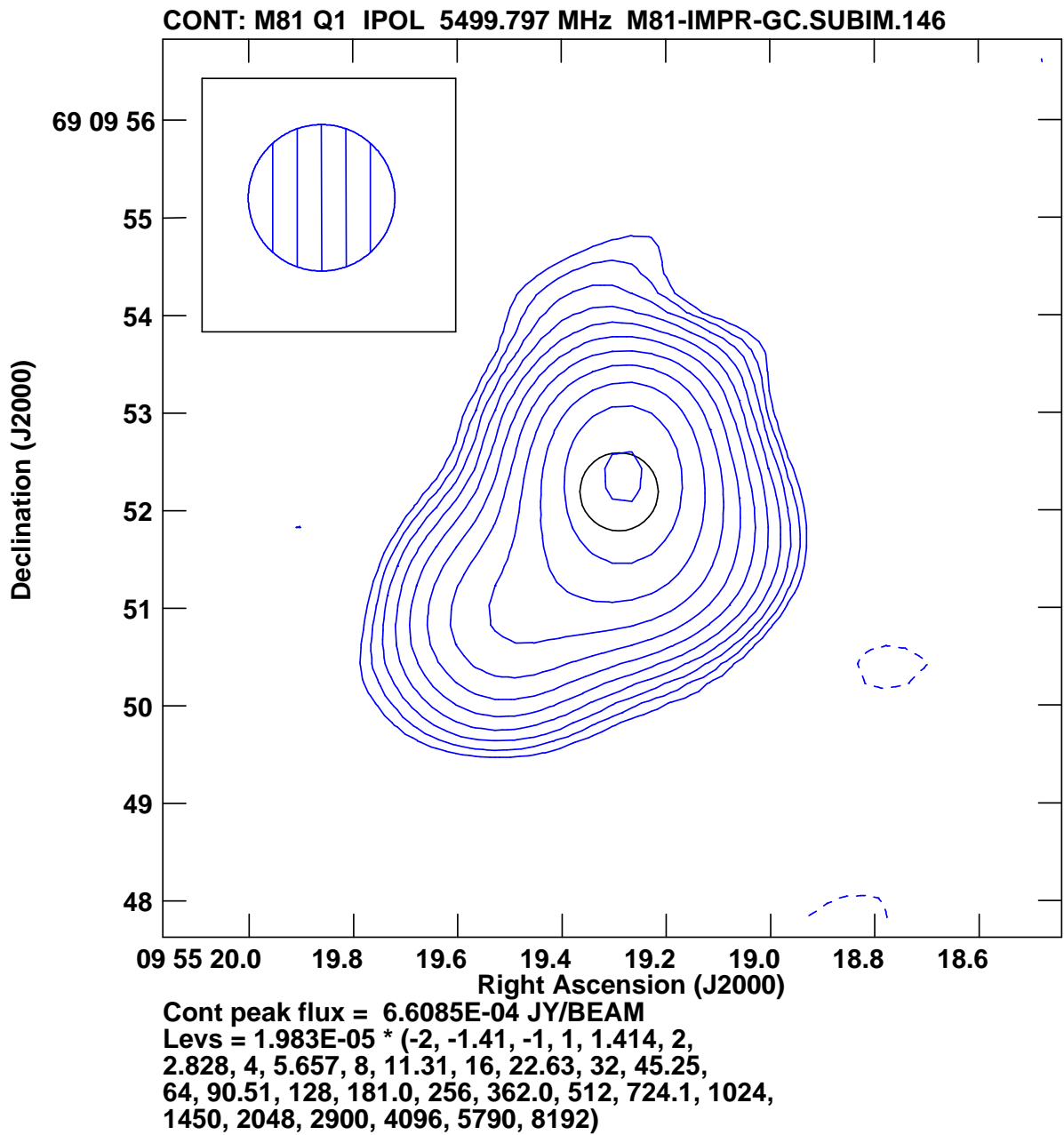}
\caption{Zoom in of the VLA cutout of the Stokes $I\/$ emission at 5.5
  GHz centered on the optical position of ID 146, an improbable GC in
  M81 (Nantais et al.\ 2011).  The diameter of the hatched circle in
  the north-east corner, 1\farcs5, is the FWHM of the VLA synthesized
  beam.  Blue contour levels are at intervals of $\sqrt{2}$ times the
  bottom contour level, which is 3\% of the image peak.  Negative
  contours are dashed and positive ones are solid.  The black central
  circle of diameter 0\farcs8 shows the optical positional uncertainty
  at 90\% confidence.  A linear scale is not given because ID 146
  might be a background galaxy.}\label{f3}
\end{figure}

\section{Implications}\label{implications}

From Figure 2, none of the 206 probable GCs in M81 is detected above
its local 3$\sigma$ level at 5.5 GHz.  The associated radio
luminosities are $\nu L_\nu = L_{\rm R} < 1.1-13 \times 10^{33}$ erg
s$^{-1}$, using a definition that implicitly assumes a flat radio
continuum spectrum up to 5.5 GHz.  Figure 4 conveys the radio
luminosities and stellar masses of each of the 206 probable GCs.  At
these radio luminosities, accreting stellar-mass compact objects in
the hard X-ray state would have radio luminosities too faint to be
detected \citep{str12b}.  What about contamination from flaring radio
emssion associated with a transition from the hard X-ray state to the
soft X-ray state?  Indeed, such flaring is thought to be the
explanation for the stellar-mass BH in M31 that achieved a peak 5-GHz
luminosity of $L_{\rm R} \sim 5.3 \times 10^{33}$ erg s$^{-1}$ and
then decayed on a timescale of days \citep{mid13,mid14}.  However, the
lack of radio detections in Figure 4 implies that such flaring
emission from stellar-mass BHs cannot be a major contaminant for these
single-epoch observations of the probable GCs in M81.  (Stronger
inferences about flares will be reported elsewhere in conjuction with
radio monitoring observations.)  This paves the way for us to
interpret Figure 4 within the context of the three approaches
mentioned in \S~\ref{motivation}, namely analogs of HLX-1
(\S~\ref{hlx}), X-ray detected clusters (\S~\ref{x-ray}), and
predictions of a semi-empirical model (\S~\ref{model}).  In the
analysis to follow, it is important to keep in mind that only 40\% of
the probable GCs have been spectroscopically confirmed \citep{nan11}.

\begin{figure}[t]
\includegraphics[angle=-90,scale=0.65]{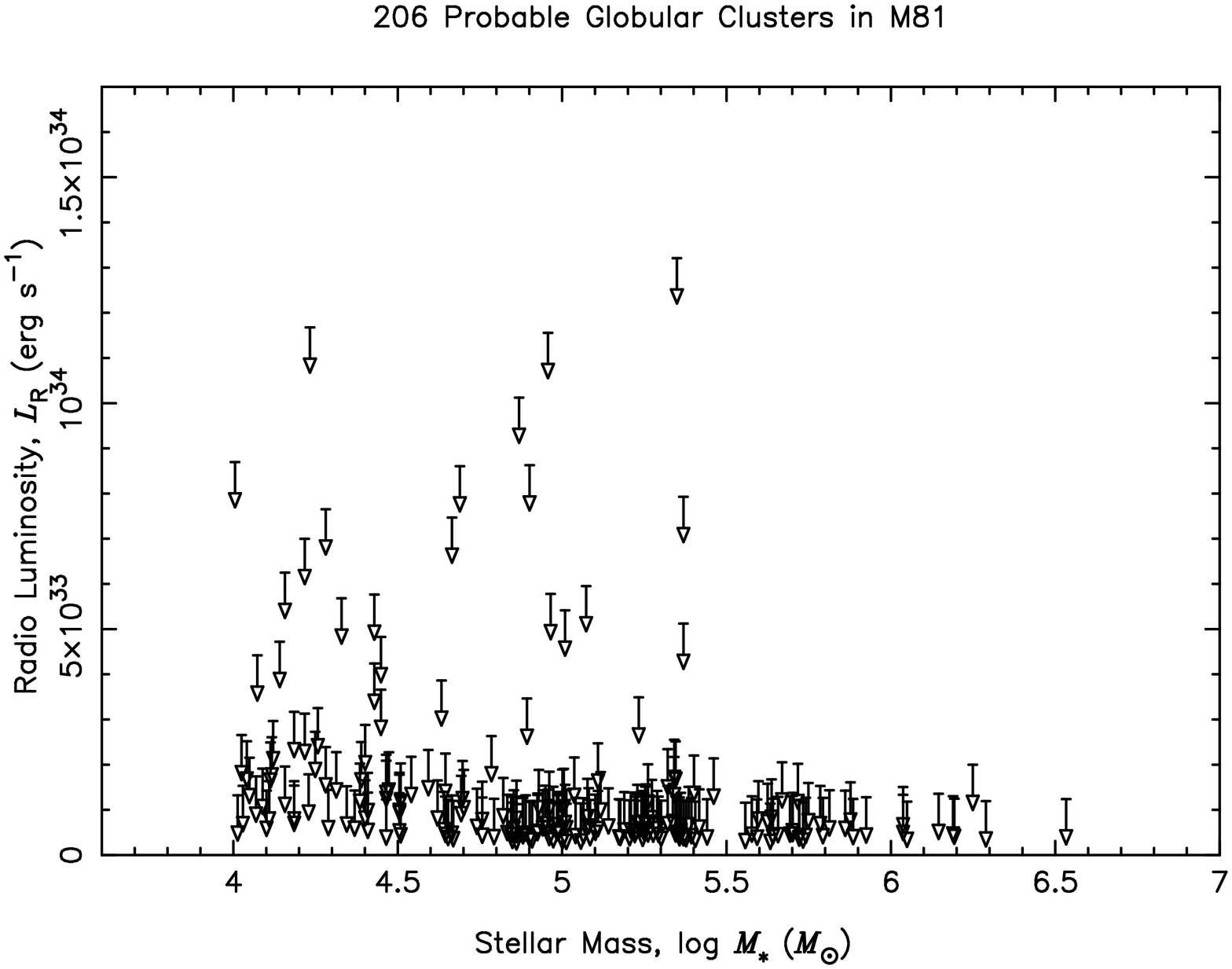}
\caption{Upper limits to the radio luminosities at 5.5 GHz, $L_{\rm
    R}$ (3$\sigma$), as a function of the stellar masses, $M_\star$,
  of 206 individual probable GCs in M81.}\label{f4}
\end{figure}

\subsection{HLX-1 Analogs}\label{hlx}

HLX-1 is a strong IMBH candidate, of mass $M_{\rm BH} \sim
10^{4-5}~M_\odot$, in an extragalactic star cluster with an observed
stellar mass of $M_\star \sim 10^{5-6}~M_\odot$
\citep{far09,sor10,wie10,far12,sor12,far14}.  The host cluster's
observed mass resembles those of the probable GCs in M81, but the age
of the host cluster's light-dominating stars, about 20 Myr, is
significantly younger than GCs, which are older than 10 Gyr
\citep{for15}.  Still, the host cluster might be more massive and
older than deduced so far, if it is actually the remnant of a stripped
dwarf galaxy \citep{far12,far14}.  As the latter authors note, such
extra mass could also help retain the gas from which the younger stars
formed.

Near 7 GHz HLX-1 can achieve a flaring luminosity of $L_{\rm R} \sim
3.4 \times 10^{36}$ erg s$^{-1}$ during transits from its hard to soft
X-ray states \citep{web12}, and a steady luminosity of $L_{\rm R} \sim
1.6 \times 10^{36}$ erg s$^{-1}$ while in its hard X-ray state
\citep{cse15}.  If that steady emission is Doppler boosted by a factor
of about five to ten, as \citet{cse15} argue, its side-on luminosity
is about $L_{\rm R} \sim 1.6-3.2 \times 10^{35}$ erg s$^{-1}$.  Given
these various radio luminosities for HLX-1, Figure 4 makes it clear
that no HLX-1 analog resides in any of the 206 probable GCs in M81.
\citet{wro15} found the same result for 337 candidate GCs in NGC\,1023
and concluded that (i) HLX-1 is accreting gas related to the formation
and/or presence of the 20-Myr-old stars in its host cluster or (ii)
the HLX-1 phenomenon is just so rare that no radio analog is expected
in NGC\,1023.  The same two conclusions apply in the case of M81.

\subsection{X-ray Detected Clusters}\label{x-ray}

\citet{sel11} used extensive {\em Chandra\/} data to localize several
hundred X-ray sources in M81, and A.\, Zezas (2015, private
communication) is investigating their matches with probable GCs.  In
the interim, we used version 1.1 of the Chandra Source Catalog
\citep{eva10} to identify X-ray sources within the optical positional
uncertainty of M81's probable GCs.  We also required that an X-ray
detection had enough significance to establish a luminosity in the
0.5-10 keV band.  This led to detections of \citet{nan11} IDs 273 and
284 as sources CXO\,J095547.0+690551 and CXO\,J095549.7+690531 with
luminosities of $L_{\rm X} \sim 9.2 \times 10^{37}$ erg s$^{-1}$ and
$L_{\rm X} \sim 4.6 \times 10^{38}$ erg s$^{-1}$, respectively.

These detections are consistent with prior {\em Chandra\/} results
\citep{swa03,liu11}.  Importantly, \citet{liu11} noted that the two
X-ray sources, detected in 16 or 17 observations, were always in a
hard X-ray state, a requisite for applying the empirical
fundamental-plane relation among the X-ray luminosity, $L_{\rm X}$,
the radio luminosity, $L_{\rm R}$, and the BH mass, $M_{\rm BH}$.  As
we wish to estimate masses, we employ the \citet{mil12} regression of
the contracted sample of \citet{plo12}.  Inserting the above values
for $L_{\rm X}$ and Figure 4's upper limits to $L_{\rm R}$ into that
regression, we estimate BH masses at 95\% confidence of $M_{\rm BH} <
99,000~M_\odot$ for ID 273 and $M_{\rm BH} < 15,000~M_\odot$ for ID
284.  The clusters' stellar masses are known from Figure 4, and lead
to estimates for the BH mass fractions of $M_{\rm BH}/M_\star < 0.18$
for ID 273 and $M_{\rm BH}/M_\star < 0.21$ for ID 284.  We comment
further on these estimates in \S~\ref{model}.  As cautioned in
\S~\ref{motivation}, we cannot rule out contamination from
X-ray-emitting stellar-mass BHs.  Indeed, that is the more likely
explanation because few IMBHs are known, whereas many X-ray binaries
are associated with extragalactic GCs.

It is also noteworthy that no probable GC in M81 has an X-ray
luminosity as high as that of HLX-1 in its hard state, $L_{\rm X} \sim
2 \times 10^{40}$ erg s$^{-1}$ \citep{god12}.  This absence of an
X-ray analog of HLX-1 is consistent with the absence of a radio analog
of HLX-1 (\S~\ref{hlx}).  However, since our limits on the radio
luminosity are over an order of magnitude lower than even a de-boosted
version of HLX-1, we are also able to rule out the presence of less
extreme sources, even in the case where such a system was heavily
absorbed and viewed edge-on, such that the X-ray emission would not be
particularly remarkable.

\subsection{Semi-Empirical Model}\label{model}

We adopt the semi-empirical model of \citet{mac08,mac10} to predict
the mass of a putative IMBH that, if experiencing hard-X-ray-state
accretion in a GC, is consistent with the upper limit on the radio
luminosity.  Following \citet{str12a} we conservatively assume that
the IMBH in a GC accretes at a fraction $f_b = 0.03$ of the Bondi rate
from a medium with a gas density $\rho = 0.2$ cm$^{-3}$.  Such values
yield a prediction for the hard-state X-ray luminosity $L_{\rm X}$.
Then, the empirical fundamental-plane regression for the \citet{plo12}
contracted sample is employed to predict the associated radio
luminosity $L_{\rm R}$.  In this way, a detection of, or upper limit
to, a radio luminosity maps to a detection of, or upper limit to, an
IMBH mass.

Figure 5 shows the results of applying this conservative,
semi-emipirical model to the 206 probable GCs in M81.  The best radio
luminosity constraint for an individual GC implies a 3$\sigma$ IMBH
mass of $M_{\rm BH} < 100,000~M_\odot$.  To reach a lower mass regime,
the AIPS task {\tt stack} was used to form a weighted-mean image stack
\citep[e.g.,][]{lin15} of the cutouts of all 206 probable GCs.  That
stack, presented in Figure 6, measures the the mean contribution to
the total radio emission from the probable GCs.  For the all-cluster
stack, the 3$\sigma$ radio-luminosity upper limit corresponds to an
IMBH mass of $\overline{M_{\rm BH}({\rm all})} < 42,000~M_\odot$.

\begin{figure}[t]
\includegraphics[angle=-90,scale=0.65]{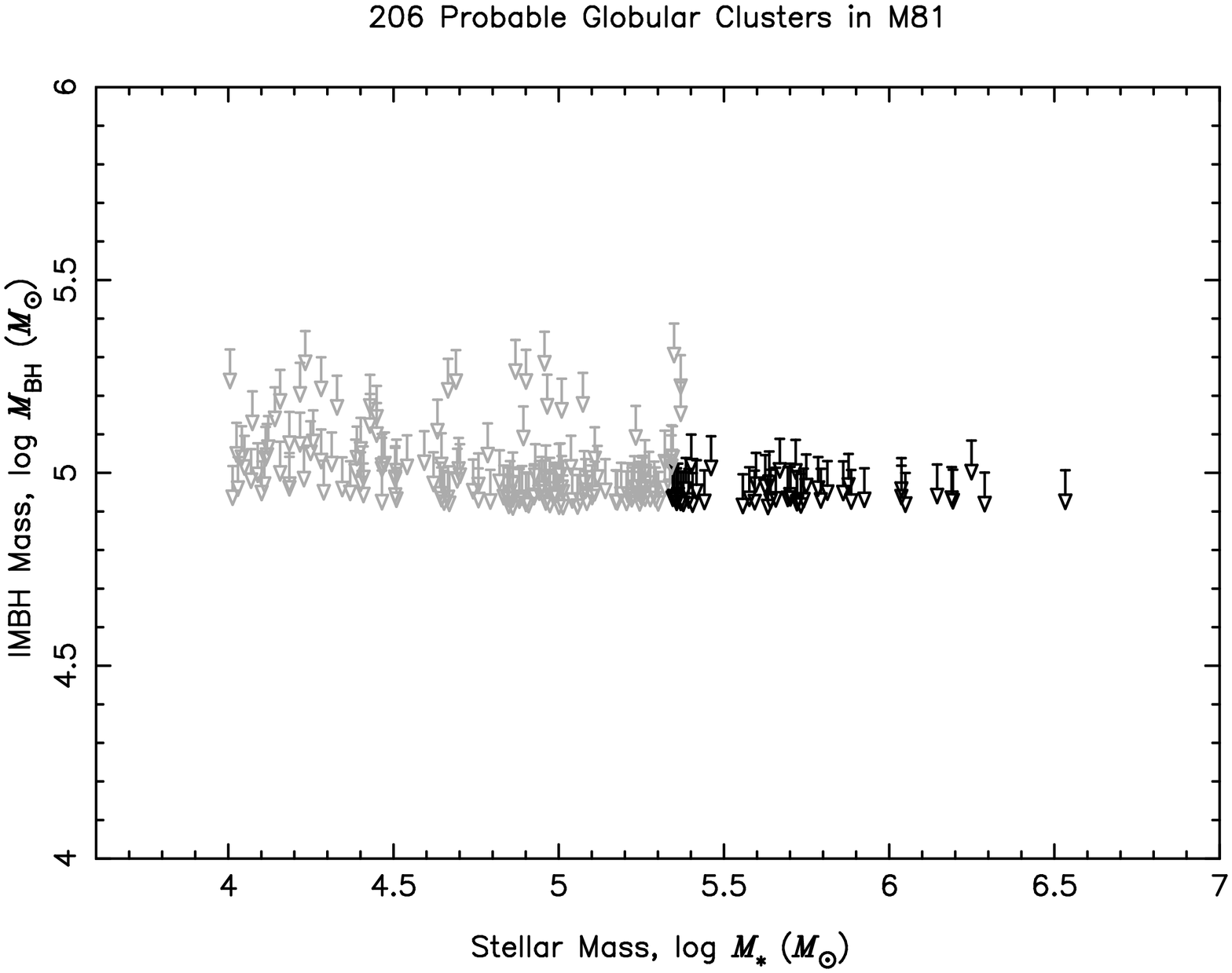}
\caption{Upper limits to the mass of the IMBH, $M_{\rm BH}$
  (3$\sigma$), as a function of the stellar masses, $M_\star$, of 206
  individual probable GCs in M81.  A semi-empirical model uses the
  radio luminosity to infer the mass of the IMBH, assuming one exists.
  The dark symbols highlight the 49 massive GCs with BH mass fractions
  $M_{\rm BH}/M_\star < 0.5$.}\label{f5}
\end{figure}

\begin{figure}[t]
\plotone{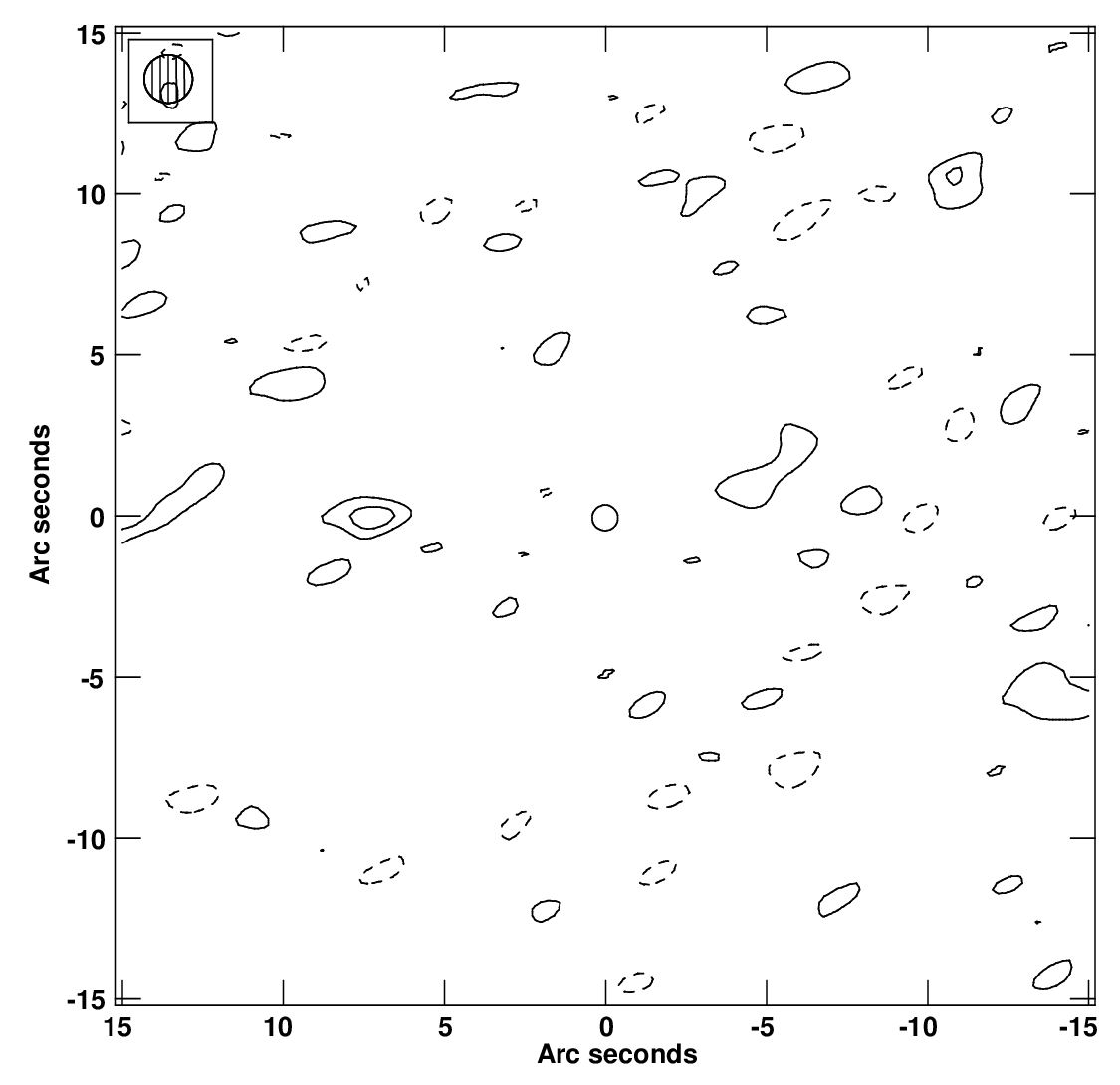}
\caption{Weighted-mean stack of the VLA images of the 206 probable
  GCs.  The stacked image has an rms noise of 0.43~$\mu$Jy~beam$^{-1}$
  (1$\sigma$).  The hatched circle, central circle and contouring
  scheme are the same as for Figure 2.  No emission is detected above
  3$\sigma$ = 1.29~$\mu$Jy~beam$^{-1}$.}\label{f6}
\end{figure}

However, the GCs' stellar masses, $M_\star$, range from about 10
thousand to 8 million solar masses, suggesting that it is physically
more relevant to focus only on the more massive GCs.  Like
\citet{wro15}, we thus focus on the GCs whose putative IMBHs have
masses that are less than half the combined mass of their stars.  The
49 GCs meeting that criterion are listed in Table 2 and plotted with
dark symbols in Figure 5.  The column contents of Table 2 are: column
1, the ID from \citet{nan11}; column 2, the stellar mass based on the
V-band (F606W) magnitudes from \citet{nan11} and that band's
mass-to-light ratio for GCs from \citet{har10}; column 3, the IMBH
mass based on the semi-empirical model; and column 4, the ratio of the
IMBH mass in column 3 to the stellar mass in column 2.  Each of the 49
massive GCs in Table 2 has a stellar mass $M_\star \gtrsim
200,000~M_\odot$.  Fully 13 of these massive GCs have upper limits to
their BH mass fractions, $M_{\rm BH}/M_\star$, in the range $0.03 -
0.15$.  Such upper limits are thus below the remarkable value of 0.15
reported for M60-UCD1 \citep{set14}.

The attributes of ID 273 in Table 2 deserve special mention: applying
the semi-empirical model leads to estimates of a BH mass of $M_{\rm
  BH} < 100,000~M_\odot$ and a BH mass fraction of $M_{\rm BH}/M_\star
< 0.19$.  For comparison, applying the empirical fundamental plane, as
in \S~\ref{x-ray}, led to estimates of $M_{\rm BH} < 99,000~M_\odot$
and $M_{\rm BH}/M_\star < 0.18$.

We also used the AIPS task {\tt stack} to form a weighted-mean image
stack of the cutouts of these 49 massive GCs.  Figure 7 shows that
stack, a measure of the the mean contribution to the total radio
emission from the 49 massive GCs.  The stack's 3$\sigma$
radio-luminosity upper limit corresponds to an IMBH mass of
$\overline{M_{\rm BH}({\rm massive})} < 51,000~M_\odot$ for M81, a
stronger constraint than the upper limit of $230,000~M_\odot$ for
NGC\,1023 \citep{wro15}.  The mean stellar mass of the massive GCs in
M81 is $\overline{M_\star({\rm massive})} = 655,000~M_\odot$.  Taking
the ratio of these values leads to a BH mass fraction
$\overline{M_{\rm BH}({\rm massive})}/\overline{M_\star({\rm
    massive})} < 0.08$ for M81, improving over the equivalent fraction
of less than 0.16 for NGC\,1023 \citep{wro15}.  The BH mass fraction
of less than 0.08 for the massive GCs in M81 is well below the BH mass
fraction of 0.15 for M60-UCD1 \citep{set14}.  Still, it should be kept
in mind that the interpretation of the BH mass fraction for M81 is
affected by unknowns like the distribution function of IMBH masses and
the fraction of massive GCs occupied by an IMBH.

\begin{figure}[t]
\plotone{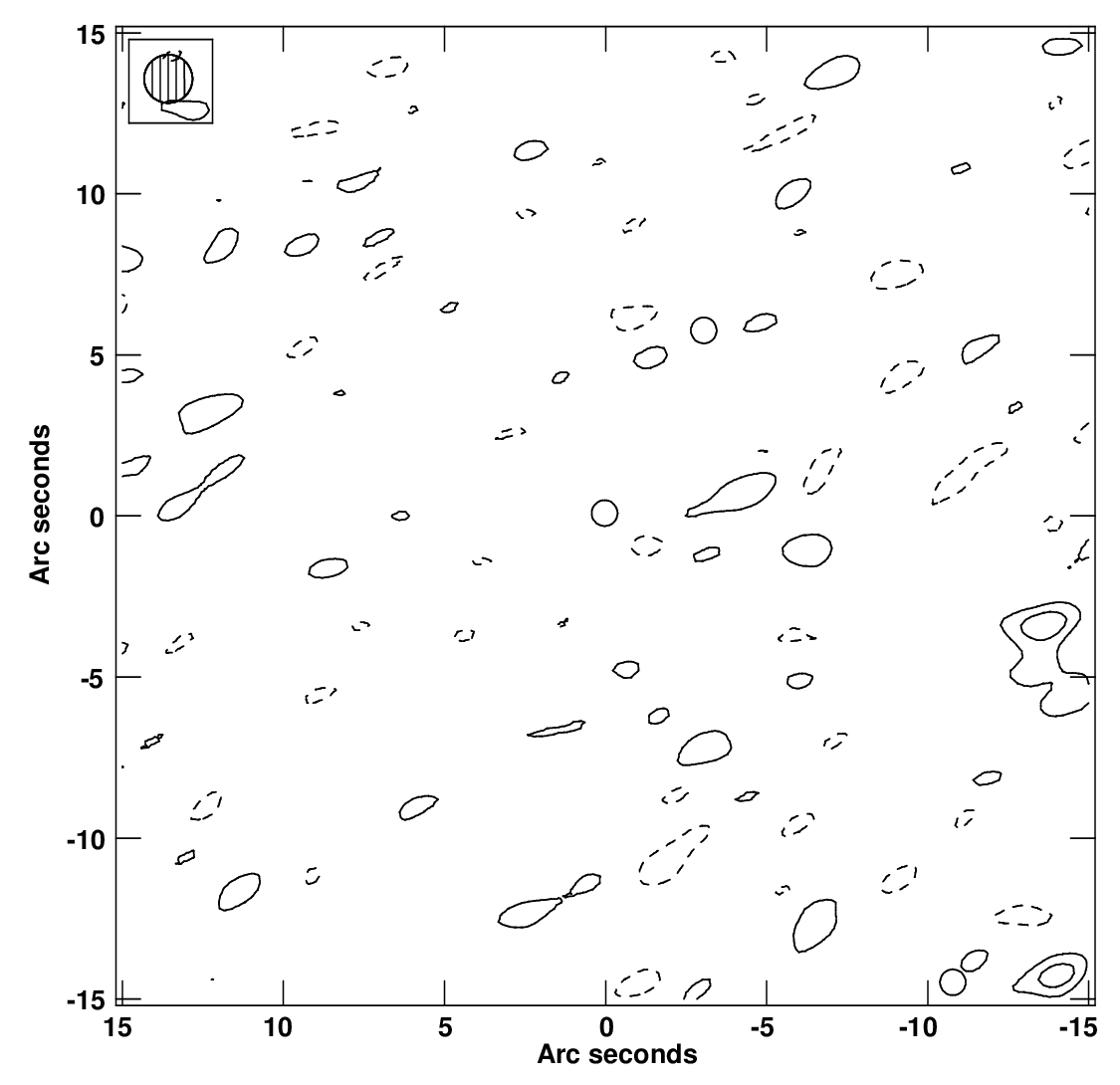}
\caption{Weighted-mean stack of the VLA images of the 49 massive GCs.
  The stacked image has an rms noise of 0.74~$\mu$Jy~beam$^{-1}$
  (1$\sigma$).  The hatched circle, central circle and contouring
  scheme are the same as for Figure 2.  No emission is detected above
  3$\sigma$ = 2.22~$\mu$Jy~beam$^{-1}$.}\label{f7}
\end{figure}

Within the context of this semi-empirical model, the VLA constraints
on individual GCs and stacks of GCs are beginning to probe the domain
of IMBHs in extragalactic GCs.  These constraints for M81 can be
improved with longer VLA exposures or similar exposures with the
next-generation VLA \citep[ngVLA;][]{car15}.  For perspective,
assuming a sufficient dynamic range given M81's LLAGN, a one-hour
exposure with the ngVLA at a frequency of 10 GHz could reach an rms
noise of 0.45~$\mu$Jy~beam$^{-1}$ for an individual GC.  Such an ngVLA
value would be very close to the rms noise achieved in Figure 6 only
after stacking the VLA images of all 206 probable GCs in M81.

\section{Summary and Conclusions}\label{sumcon}

We used a four-pointing VLA mosaic at 5.5 GHz to search for the
radiative signatures of IMBH accretion from 206 probable GCs in M81, a
spiral galaxy at a distance of 3.63 Mpc.  None of the individual GCs
were detected.  Similarly, only upper limits were obtained from
weighted-mean image stacks of all 206 GCs and of the 49 massive GCs
with $M_\star \gtrsim 200,000~M_\odot$.  We examined the implications
of these data for IMBHs, if any exist in these GCs.  Our principal
findings are as follows:

\begin{enumerate}
\item The 206 GCs in M81 lack radio analogs of HLX-1, a strong IMBH
  candidate in a star cluster in the early-type galaxy ESO\,243-49.
  This suggests that HLX-1 is accreting gas related to the 20-Myr-old
  stars in its host cluster or that the HLX-1 phenomenon is so rare
  that no radio analog is expected in M81.

\item Two GCs exhibit hard-state X-ray emission. From the empirical
  fundamental-plane relation, their X-ray and radio luminosities
  suggest individual IMBH masses, $M_{\rm BH}$, of less than
  $99,000~M_\odot$ and $15,000~M_\odot$, and associated BH mass
  fractions, $M_{\rm BH}/M_\star$, of less than 0.18 and 0.21.  With
  only upper limits to the radio luminosities, we cannot rule out the
  likely scenario of contamination from X-ray-emitting stellar-mass
  BHs in these GCs.

\item A semi-empirical model developed for Milky Way GCs converts the
  upper limits on radio luminosities to upper limits on IMBH masses.
  Applying this model to M81, over a dozen individual GCs appear to
  have upper limits on the BH mass fractions, $M_{\rm BH}/M_\star$,
  that are below the noteworthy value of 0.15 reported for M60-UCD1.
  Also, the M81 stacks correspond to IMBH masses of $\overline{M_{\rm
      BH}({\rm all})} < 42,000~M_\odot$ for all the GCs and to
  $\overline{M_{\rm BH}({\rm massive})} < 51,000~M_\odot$ for the
  massive GCs.  This model is making inroads into the
  difficult-to-observe regime of IMBHs in extragalactic GCs.
\end{enumerate}

\acknowledgments We thank the referee for a helpful and timely report,
and Dr.~E. Greisen for providing the new AIPS task {\tt stack}.  JCAMJ
is the recipient of an Australian Research Council Future Fellowship
(FT140101082).  This research has made use of data obtained from the
Chandra Source Catalog, provided by the Chandra X-ray Center (CXC) as
part of the Chandra Data Archive.  This research has made use of the
NASA/IPAC Extragalactic Database (NED) which is operated by the Jet
Propulsion Laboratory, California Institute of Technology, under
contract with the National Aeronautics and Space Administration.  Any
opinions, findings, and conclusions or recommendations expressed in
this material are those of the authors and do not necessarily reflect
the views of the National Science Foundation.

{\it Facilities:} \facility{CXO, VLA}.


\clearpage
\begin{deluxetable}{rrrr}
\tablecolumns{4}
\tablewidth{0pc}
\tablecaption{VLA Pointing Centers and Observation Dates}
\tablehead{
\colhead{R.A.}    & \colhead{Decl.}   &
\colhead{UT Date} & \colhead{MJD}     \\
\colhead{(J2000)} & \colhead{(J2000)} & 
\colhead{}        & \colhead{}}
\startdata
09 55 40.71 & 69 07 19.45 & 2014 Jan 5 & 56662.15$\pm$0.02 \\
09 54 58.45 & 69 04 43.90 & 2014 Jan 6 & 56663.11$\pm$0.02 \\
09 56 09.55 & 69 03 20.83 & 2014 Jan 5 & 56662.12$\pm$0.02 \\
09 55 28.20 & 69 00 51.34 & 2014 Jan 6 & 56663.16$\pm$0.02 \\
\enddata
\tablecomments{Units of right ascension are hours, minutes, and
  seconds, and units of declination are degrees, arcminutes, and
  arcseconds.}
\end{deluxetable}

\clearpage
\begin{deluxetable}{rrrr}
\tablecolumns{4}
\tablewidth{0pc}
\tablecaption{Massive Globular Clusters in M81}
\tablehead{
\colhead{ID} & \colhead{$M_\star$} & 
\colhead{$M_{\rm BH}$} & \colhead{$M_{\rm BH}$/$M_\star$} \\
\colhead{} & 
\colhead{($M_\odot$)} & \colhead{($M_\odot$)} & \colhead{}}
\startdata
  28 & 0.52E+06 & $<$0.12E+06 & $<$0.23 \\
  31 & 0.43E+06 & $<$0.11E+06 & $<$0.26 \\
  71 & 0.26E+06 & $<$0.11E+06 & $<$0.41 \\
  90 & 0.11E+07 & $<$0.10E+06 & $<$0.10 \\
 100 & 0.28E+06 & $<$0.10E+06 & $<$0.37 \\
 115 & 0.39E+06 & $<$0.10E+06 & $<$0.26 \\
 116 & 0.44E+06 & $<$0.10E+06 & $<$0.23 \\
 118 & 0.19E+07 & $<$0.10E+06 & $<$0.05 \\
 136 & 0.23E+06 & $<$0.10E+06 & $<$0.44 \\
 145 & 0.25E+06 & $<$0.10E+06 & $<$0.39 \\
 158 & 0.53E+06 & $<$0.10E+06 & $<$0.19 \\
 160 & 0.34E+07 & $<$0.10E+06 & $<$0.03 \\
 162 & 0.11E+07 & $<$0.10E+06 & $<$0.09 \\
 175 & 0.45E+06 & $<$0.10E+06 & $<$0.23 \\
 190 & 0.53E+06 & $<$0.10E+06 & $<$0.19 \\
 199 & 0.11E+07 & $<$0.11E+06 & $<$0.10 \\
 201 & 0.23E+06 & $<$0.10E+06 & $<$0.43 \\
 209 & 0.62E+06 & $<$0.10E+06 & $<$0.16 \\
 218 & 0.47E+06 & $<$0.12E+06 & $<$0.26 \\
 226 & 0.22E+06 & $<$0.10E+06 & $<$0.47 \\
 227 & 0.49E+06 & $<$0.10E+06 & $<$0.21 \\
 228 & 0.84E+06 & $<$0.10E+06 & $<$0.12 \\
 232 & 0.65E+06 & $<$0.11E+06 & $<$0.17 \\
 236 & 0.22E+06 & $<$0.10E+06 & $<$0.47 \\
 239 & 0.25E+06 & $<$0.10E+06 & $<$0.41 \\
 246 & 0.15E+07 & $<$0.10E+06 & $<$0.07 \\
 247 & 0.24E+06 & $<$0.10E+06 & $<$0.42 \\
 256 & 0.36E+06 & $<$0.99E+05 & $<$0.28 \\
 269 & 0.16E+07 & $<$0.10E+06 & $<$0.07 \\
 273 & 0.55E+06 & $<$0.10E+06 & $<$0.19 \\
 275 & 0.50E+06 & $<$0.10E+06 & $<$0.21 \\
 277 & 0.43E+06 & $<$0.99E+05 & $<$0.23 \\
 282 & 0.50E+06 & $<$0.11E+06 & $<$0.21 \\
 288 & 0.29E+06 & $<$0.12E+06 & $<$0.43 \\
 292 & 0.54E+06 & $<$0.99E+05 & $<$0.18 \\
 293 & 0.61E+06 & $<$0.11E+06 & $<$0.18 \\
 294 & 0.75E+06 & $<$0.11E+06 & $<$0.15 \\
 295 & 0.23E+06 & $<$0.11E+06 & $<$0.46 \\
 301 & 0.77E+06 & $<$0.10E+06 & $<$0.13 \\
 302 & 0.73E+06 & $<$0.11E+06 & $<$0.15 \\
 304 & 0.23E+06 & $<$0.10E+06 & $<$0.45 \\
 307 & 0.40E+06 & $<$0.11E+06 & $<$0.28 \\
 315 & 0.38E+06 & $<$0.10E+06 & $<$0.27 \\
 330 & 0.24E+06 & $<$0.11E+06 & $<$0.45 \\
 340 & 0.42E+06 & $<$0.11E+06 & $<$0.26 \\
 351 & 0.14E+07 & $<$0.11E+06 & $<$0.08 \\
 388 & 0.56E+06 & $<$0.11E+06 & $<$0.20 \\
 398 & 0.18E+07 & $<$0.12E+06 & $<$0.07 \\
 410 & 0.25E+06 & $<$0.13E+06 & $<$0.50 \\
\enddata
\end{deluxetable}

\end{document}